\documentclass[amsmath,aps,10pt,twocolumn,superscriptaddress]{revtex4} 

\usepackage{amssymb,url}

\usepackage{subfigure}
\usepackage{graphicx}
\usepackage{hyperref}

\def\beq{\begin{equation}}
\def\eeq{\end{equation}}

\def\bea{\arraycolsep .1em \begin{eqnarray}}
\def\eea{\end{eqnarray}}

\def\eq#1{(\ref{#1})}

\def\s0#1#2{\mbox{\small{$ \frac{#1}{#2} $}}}
\def\0#1#2{\frac{#1}{#2}}

\def\grgl{\:\hbox to -0.2pt{\lower2.5pt\hbox{$\sim$}\hss}{\raise3pt\hbox{$>$}}\:}
\def\klgl{\:\hbox to -0.2pt{\lower2.5pt\hbox{$\sim$}\hss}{\raise3pt\hbox{$<$}}\:}



\begin{document}

\title{Limit cycles and quantum gravity}

\author{Daniel Litim}
\affiliation{Department of Physics \& Astronomy, 
               University of Sussex, Brighton BN1 9QH, UK}
\author{Alejandro Satz}
\affiliation{
Center for Fundamental Physics, University of Maryland, College Park, MD 20742-4111, USA} 

\begin{abstract}
We study renormalization group equations 
 of quantum gravity in four dimensions. We find an ultraviolet fixed point in accordance with the asymptotic safety conjecture, and infrared fixed points corresponding to general relativity with positive, vanishing or negative cosmological constant.  In a minisuperspace approximation, we additionally find a renormalization group limit cycle shielding the ultraviolet from the infrared fixed points. We discuss implications of this pattern for asymptotically safe gravity in the continuum and on the lattice.
\end{abstract}

\maketitle

 Renormalization group (RG) ideas have had a deep impact on the understanding of  quantum field theory and statistical physics \cite{Wilson:1993dy}. They give access to the scale- or energy-dependence of couplings in many areas of physics including condensed matter, nuclear and particle physics, and gravitation. 
Fixed points are the simplest asymptotic solutions of the RG and describe  systems whose couplings no longer change with scale, leading to scale invariance and universal scaling laws. 
Infrared (IR) fixed points  often govern the long-distance behavior, 
whereas ultraviolet (UV) fixed points control the short-distance structure, central for a fundamental definition of quantum field theory and, possibly, gravity.

Limit cycles offer an intriguing, and more complex, possibility 
for the  asymptotic behavior of the RG
 \cite{Wilson:1970ag}. 
  They arise if couplings 
    change with scale in a cyclic manner leading to log-periodicity
 and discrete scale invariance. 
 Unlike fixed points, limit cycles have been a rare encounter in quantum field theory, and finding them is more challenging as it requires global rather than local information about the RG. 
 Recent studies of limit cycles in particle physics models, however,  
suggests  that their presence could be much more commonplace  \cite{Fortin:2011ks}.  Further systems with known limit cycles include
effective theories for bound states  \cite{Efimov:1970zz}, QCD with quark masses  slightly different from the physical  values \cite{Braaten:2003eu}, and 
discrete Hamiltonians \cite{Glazek:2002hq}. 
Discrete scale invariance  also occurs in  contexts as diverse as turbulence, earthquakes, and stock market crashes 
\cite{Sornette}, or gravitational collapse \cite{Choptuik:1992jv}.

In quantum gravity, fixed point studies have received substantial attention in the past decade \cite{Litim:2006dx} (see \cite{Litim:2011cp} for recent overviews).
The existence of an UV fixed point  offers the intriguing possibility that metric gravity becomes a well-defined quantum theory alongside the other fundamentally known forces  \cite{Weinberg}. By now, UV fixed points have been found using RG  methods in the continuum \cite{Reuter:1996cp,Souma:1999at,Lauscher:2001ya,Litim:2003vp,Bonanno:2004sy,Niedermaier:2009zz} and numerical simulations on the lattice \cite{Hamber:1999nu,Ambjorn:2011cg}.  
Equally important is the existence of well-defined RG trajectories which  connect the UV fixed point  at short distances with classical general relativity at large distances. This link has been established  both in four-dimensional gravity \cite{Reuter:2001ag} and higher dimensional  extensions \cite{Litim:2003vp,Fischer:2006fz}, thus providing explicit realizations of S.~Weinberg's asymptotic safety conjecture \cite{Weinberg}.

It is the purpose of this Letter to show that gravity may also display a limit cycle under its renormalization group flow.  Our observations derive from simplified models of $4d$ quantum gravity which include Newton's coupling and the cosmological constant as running parameters. 
We discuss the origin and properties of the gravitational limit cycle, and its fate in the physical theory. Some implications for  the  asymptotic safety conjecture 
in the continuum
and on the lattice are also indicated.
To achieve our picture, we 
study euclidean Einstein-Hilbert gravity in dynamically different settings  
starting with the classical 
action in four dimensions, 
\begin{equation}
\label{EH}
S=\int d^4x\sqrt{{\rm det} g_{\mu\nu}}
\frac{1}{16\pi\,G}
\left[-R+2\Lambda\right]\,.
\end{equation}
Here,  $G= 6.67\times10^{-11} {\rm m}^3/({\rm kg}\,{\rm s}^2)$ denotes Newton's coupling, $\Lambda$ the cosmological constant, and $R$ the Ricci scalar. The metric field $g_{\mu\nu}$ is the fundamental degree of freedom.
We also study a minisuperspace approximation \cite{misner69}
with spatially flat Friedmann-Robertson-Walker (FRW) metric
$ds^2=a(t)^2[N^2(t)dt^2+dr^2+r^2d\Omega^2]$.
The  coordinate $t$ corresponds to the Wick-rotated version of the FRW  conformal time parameter once the lapse function $N(t)$ is gauged to a constant $(N=1)$, leading to
\begin{equation}
\label{S_MS}
S= \int dt\,\frac{3v}{8\pi G}\left[-a'(t)^2+\frac{\Lambda}{3}a(t)^4\right]\,.
\end{equation}
Here $v$ denotes the spatial 3-volume, and
the dynamics of gravity is reduced to the  scale factor $a(t)$. The negative sign of the kinetic term hints at the unboundedness of the conformal mode in the classical theory. A closely related approximation is  the conformally reduced model \cite{Reuter:2008wj}
 \begin{equation}
\label{S_CR}
S=\int d^4x\sqrt{\hat g}\,
\frac{3}{8\pi G}
\left[\chi\hat\Box\chi-\frac{\hat R}{6}\chi^2+\frac{\Lambda}{3}\chi^4
\right].
\end{equation}
The conformal factor  $\chi(x)$  is linked to the original metric field via $g_{\mu\nu}(x)=\chi(x)^2\hat g_{\mu\nu}(x)$, and the Laplacean and the Ricci scalar are now functions of a fixed reference metric $\hat g_{\mu\nu}(x)$ \cite{Reuter:2008wj}.  
In comparison with \eq{EH}, the actions  \eq{S_MS} and \eq{S_CR}  neglect spin-two modes,
and the conformal factor in \eq{S_MS} only depends on euclidean time,
rather than euclidean space-time \eq{S_CR}.

We now employ the RG to  analyse the differences between  \eq{EH}, \eq{S_MS} and \eq{S_CR}  on the quantum level.
 Fluctuations turn the couplings $G$ and $\Lambda$  into running couplings $G_k$ and $\Lambda_k$ as a function of the RG momentum or energy scale $k$. At the same time the actions \eq{EH}, \eq{S_MS} and \eq{S_CR} 
become   `flowing' effective action $S\to \Gamma_k$.  Following the ideas by L.~Kadanoff and K.~Wilson, RG equations which control the scale-dependence of couplings are obtained by introducing a momentum cutoff into the theory.
The change of the flowing action with the cutoff scale is expressed by an exact functional identity \cite{Wetterich:1992yh}
\begin{equation}
\label{FRG}
\partial_k\Gamma_k=\frac12 {\rm Tr}\left({\Gamma^{(2)}_k+R_k}\right)^{-1}\partial_kR_k\,.
\end{equation}
The trace denotes a sum over all propagating  modes, $\Gamma^{(2)}_k$ stands for the second functional derivative, and $R_k$ for a properly chosen Wilsonian momentum cutoff at the energy scale $k$ \cite{Litim:2001up}. The physics meaning of  \eq{FRG} is that the fluctuations of modes at energy scale $k$ drive the RG running  at that scale. The RG  flow \eq{FRG} serves as a master equation for the running of all couplings between the UV ($k\to\infty)$ and IR $(k\to 0)$ limits of the theory.

For the theories given by \eq{EH} and \eq{S_CR}, the RG  flow  \eq{FRG}   has been obtained in \cite{Reuter:1996cp,Lauscher:2001ya,Litim:2003vp,Bonanno:2004sy,Codello:2008vh,Reuter:2008wj}, also using \cite{Litim:2001up,Litim:2001dt} and the background field formalism. The RG flow for \eq{S_MS} resembles that of a quantum mechanical system because the momentum trace in \eq{FRG} becomes lower-dimensional \cite{LitimSatz2012}. 
In all three cases, and using dimensionless couplings $g\equiv G_k\,k^{2}$ and $\lambda\equiv \Lambda_k/k^2$, the RG running of couplings takes the simple form 
\begin{equation}
\label{beta}
\begin{array}{rcl}
\displaystyle
\beta_g
&=&
\displaystyle
\left( 2+\eta\right)\,g\,,\quad\quad\eta=-g\,a_1\\[1ex]
\beta_\lambda
&=&
\displaystyle
-(2-\eta)\lambda+g\,a_2-\frac{g\,a_3}{2+n}\,\eta
\end{array}
\end{equation}
where $\beta_g\equiv\frac{{\rm d}g}{{\rm d}\ln k}$ (idem for $\lambda$). The fluctuation-induced functions $a_1$, $a_2$ and $a_3$  and the parameter $n$ depend on the model.
Throughout, the graviton anomalous dimension $\eta=\frac{{\rm d}\ln G_k}{{\rm d}\ln k}$  is approximated by its leading term in $g$.  
It vanishes  in the absence of fluctuations where \eq{beta} reduce to $\beta_g=2g$ and $\beta_\lambda=-2\lambda$. This is the semi-classical regime where $G_k$ and $\Lambda_k$ are constants. At a non-trivial fixed point $\eta$ takes the value $-2$ leading to a weak (strong) gravity regime in the UV (IR)  \cite{Litim:2006dx}. 
Higher order corrections such as Hartree-Fock type resummations for $\eta$ can be studied as well \cite{Reuter:1996cp,Lauscher:2001ya,Litim:2003vp,Codello:2008vh}.

\begin{figure}[t]
\begin{center}
\unitlength0.001\hsize
\begin{picture}(1000,630)
\put(20,0){\includegraphics[scale=.75]{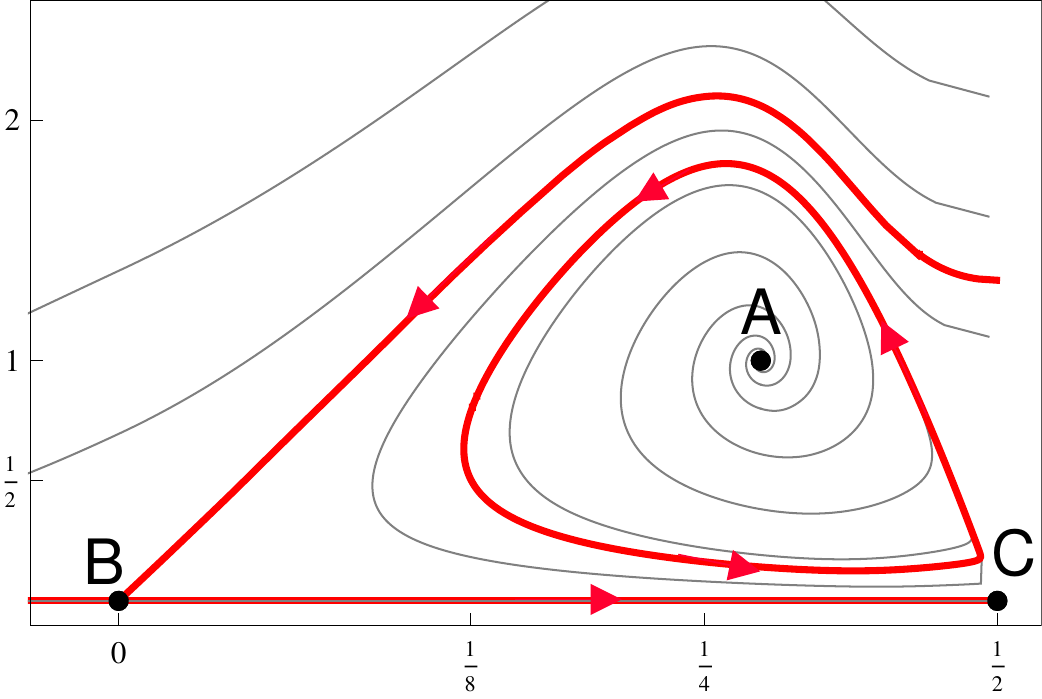}}
\put(-20,570){\Large$\frac{g}{g_*}$}
\put(940,10){\Large $\lambda$}
\end{picture}
\caption{Fixed points and limit cycle in the minisuperspace approximation (in all figures  arrows point 
to the infrared 
and the lower axes are scaled as $\lambda\to\frac{\lambda}{1+2|\lambda|}$  for display purposes).}
\vskip-1cm
\label{pMS}
\end{center}\end{figure}

We begin with the phase diagram of the minisuperspace. The definition of $g$ is  fixed up to an overall normalization, linked to the freedom in fixing the 3-volume \cite{LitimSatz2012}. 
With a suitable normalization we find \eq{beta} with
\begin{equation}
\label{aMS}
\begin{array}{rcl}
a_1&=&
\displaystyle
\frac{2}{3\pi} 
\frac{\lambda^2}{(1-2\lambda)^4}
\,,\quad
a_{2}=
a_{3}=
\displaystyle
\frac{1}{4\pi}
\frac{1}{1-2\lambda}
\end{array}
\end{equation}
 and $n=1$. Note that the RG flow for Newton's coupling is only driven by the cosmological constant as $a_1$ vanishes for vanishing $\lambda$. We find four distinguished fixed points which we denote by $A, B, C$ and $D$ (see Fig.~\ref{pMS}): The point $A$ stands for an UV stable fixed point with coordinates $(g,\lambda)_*=(3.68,0.288)$. The  universal eigenvalues  are a complex conjugate pair of exponents $\theta=1.77 \pm 7.31 i$,
 and the positive sign of the real part indicates UV stability in both couplings. 
At this fixed point the theory becomes asymptotically safe \cite{Weinberg}.  
In its vicinity, we have that
\beq\label{LC}
\lambda_*-\lambda(t+T)=e^{-D}(\lambda_*-\lambda(t))
\eeq 
and the same for $g$, where $t=\ln k$  is the logarithmic RG time,  $T=\frac{2\pi}{|{\rm Im}\,\theta|}$  and $D={\rm Re}\,\theta$.
The behavior \eq{LC} resembles that of an UV limit cycle whose radius shrinks by $e^{-D}\approx 0.17$ after each cycle with periodicity $T\approx0.86$ about the nodal point $(\lambda_*,g_*)$.
We also have a Gaussian infrared fixed point  $(g,\lambda)_*=(0,0)$ at vanishing coupling denoted by $B$ with scaling exponents $\theta=\pm2$,
and a degenerate IR fixed point  $C$ at  $(g,\lambda)=(0,\frac12)$ \cite{inprep}. Finally, there is  an IR attractive fixed point $D$, located at negative cosmological constant  $(g,\frac1{\lambda})_*=(0,0^-)$ with two infrared attractive scaling exponents $\theta=-2$ (not displayed).  Along the line $(g\neq0,\lambda=\frac12)$, we have $1/\eta=0$ indicating the limit of validity for our approximation. 

   \begin{figure*}[t]
\begin{center}
\unitlength0.001\hsize
\begin{picture}(1000,310)
\put(60,0){\includegraphics[scale=.75]{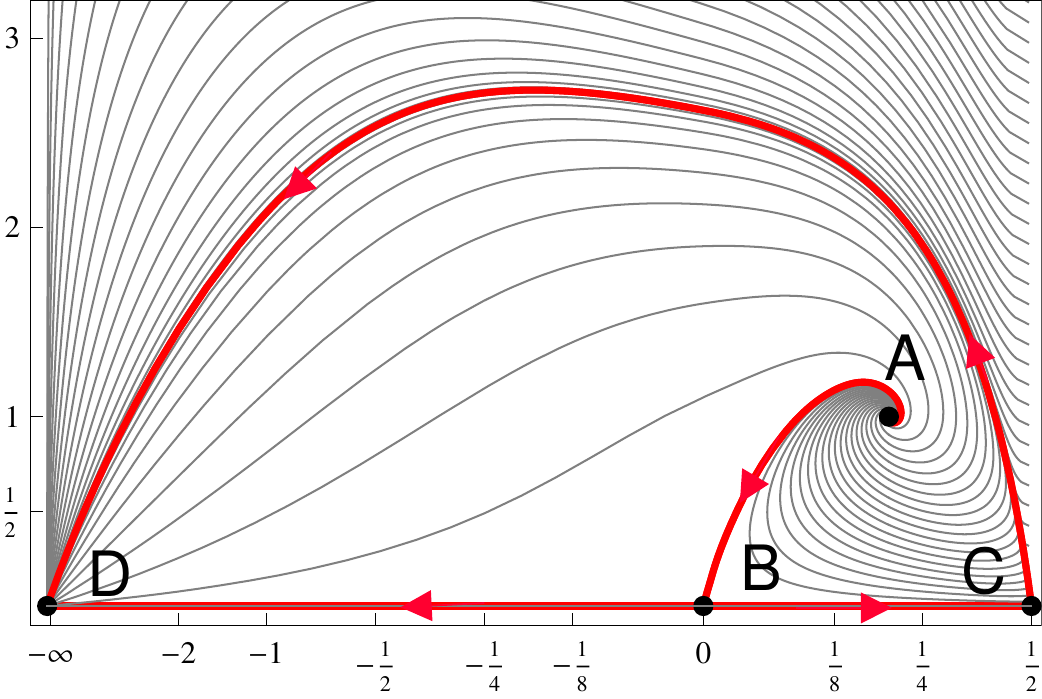}}
\put(520,0){\includegraphics[scale=.75]{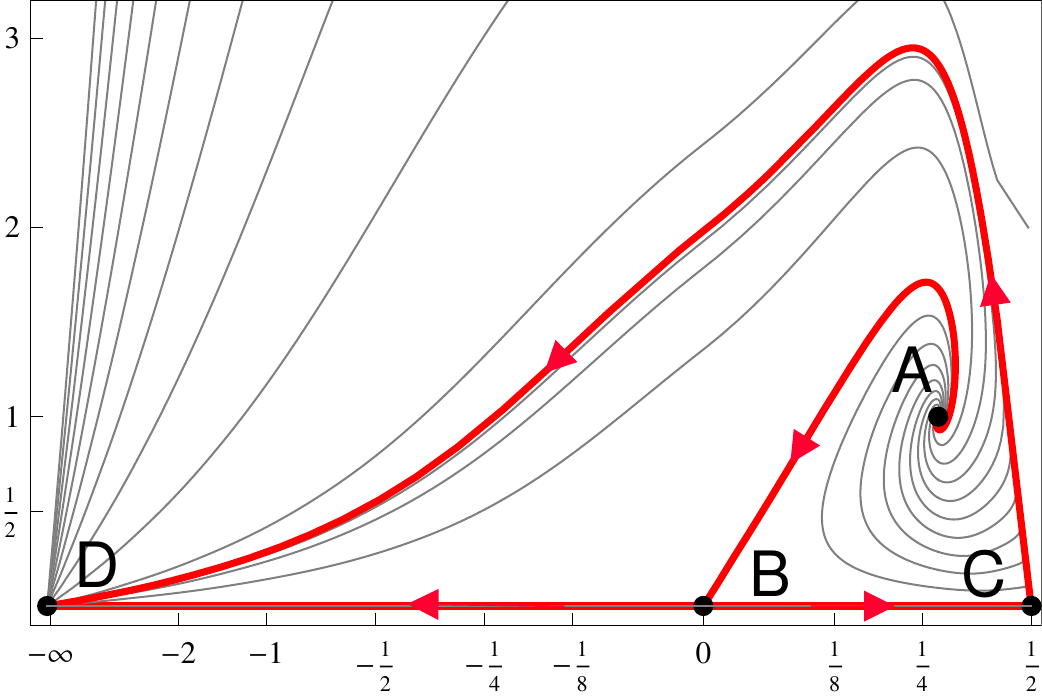}}
\put(25,280){\Large$\frac{g}{g_*}$}
\put(505,-10){\Large $\lambda$}
\end{picture}
\caption{Fixed points and separatrices in the Einstein-Hilbert  approximation (left) and the conformally reduced model (right).}
 \vskip-.7cm
\label{p4d}
\end{center}\end{figure*}

The phase diagram is split into three  parts. The most interesting feature is the appearance of an IR limit cycle (thick line) which attracts all
trajectories emanating from the UV fixed point $A$. The limit cycle implies log-periodic behavior  $g(t)= g(t+T)$ and $\lambda(t)= \lambda(t+T)$ corresponding to \eq{LC} with $D=0$ and periodicity 
$T\approx1.571$. Flow lines emanating from the line $(g,\lambda\approx\frac12)$ terminate either  at the limit cycle, 
at the Gaussian fixed point $B$ (thick line), or at the fixed point $D$, depending on whether $g<g_c\approx 1.33 g_*$, $g=g_c$ or $g>g_c$.
An extended semi-classical regime is achieved for trajectories starting just below the separatrix (thick line), at $(g,\lambda\approx\frac12)$ and $g<g_c$, as these can spend substantial RG time in the vicinity of $B$ before terminating in the limit cycle. Such trajectories 
 imply modifications of gravity in the deep IR.
On the limit cycle, the dimensionful couplings $G_k$ and $1/\Lambda_k$ grow by the universal factor $e^{2T}\approx 23.16$ after each cycle, leading to  strong gravity $G_k\gg G$ and a vanishing cosmological constant $\Lambda_k\to 0$ in the IR limit.
The average growth rate  $\langle G(t)/G(t_0)\rangle_T = (k_0/k)^2$
where $t-t_0$ is an integer multiple of $T$, is exactly the same as that for a non-trivial IR fixed point.
The cycle average $\langle g(t)\lambda(t)\rangle_T\approx0.455\, g_*\lambda_*$ is smaller than its value at the UV fixed point.
None of these trajectories, however,  is UV safe.
Those which are, $i.e.$~all trajectories emanating from the UV safe fixed point, display a cross-over into the IR limit cycle too narrow in logarithmic RG time to allow for an extended semi-classical regime. We conclude that the minisuperspace approximation \eq{S_MS} with \eq{beta}, \eq{aMS}
does not admit
trajectories connecting the UV with the IR as required by the asymptotic safety conjecture, despite the presence of the relevant UV and IR fixed points.

Next we enhance the dynamical content and turn to Einstein-Hilbert gravity \eq{EH} in terms of  the metric field $g_{\mu\nu}$ and ghosts
in Landau-de~Witt  gauge \cite{Lauscher:2001ya,Litim:2003vp}. 
As opposed to \eq{aMS}, the RG flow receives extra contributions from self-interactions such as spin two, the ghosts, and the conformal mode.   One finds \eq{beta} with
\begin{equation}\begin{array}{rcl}
a_1&=&
\displaystyle
\frac{81-132\lambda+100\lambda^2}{24\pi(1-2\lambda)^2}
\\[2ex]
a_2&=&
\displaystyle
\frac{1}{2\pi} \frac{1+4\lambda}{1-2\lambda}\,,\quad
a_3=
\frac{1}{2\pi}\frac{5-4\lambda}{1-2\lambda}
\end{array}
\label{Landau}
\end{equation}
and $n=4$. 
Interestingly, we again find the same
 fixed points $A, B, C$ and $D$ (see Fig.~\ref{p4d}, left panel),
 the only difference being that the coordinates $(g,\lambda)_*=(0.940,0.197)$ have changed. The  eigenvalues remain
  a complex conjugate pair of exponents \cite{Souma:1999at,Lauscher:2001ya,Litim:2003vp,Niedermaier:2009zz} and, here, read $\theta=1.49 \pm 2.68 i$.
On the other hand, the topology of the RG flow has changed substantially.
The UV fixed point is now connected with the Gaussian one by a separatrix $AB$ (thick line). A second separatrix appears connecting the IR fixed points $C$ and $D$  (thick line). It also acts as an infrared attractor for all (but one) trajectories emanating from the UV fixed point (thin lines), which circulate counter-clockwise into the separatrix $CD$. The latter is thus 
a remnant of the minisuperspace's limit cycle.  Ultimately all trajectories are attracted towards the weak-gravity regime dictated by the IR fixed points $C$ and $D$. This way the RG flow shields the theory from the strong gravity regime 
observed in the minisuperspace.
The phase diagram also admits asymptotically safe RG trajectories connecting the UV fixed point with an extended semi-classical regime with either positive, vanishing, or negative cosmological constant. This picture  
should be compared with the phase diagram of conformally reduced gravity \eq{S_CR} with RG flow  \eq{beta}, \eq{aMS} and $n=4$ \cite{Reuter:2008wj}. We find the same  fixed points as discussed above (Fig.~\ref{p4d}, right panel), except that the coordinate of the UV fixed point has changed to $(g,\lambda)_*=(4.65, 0.279)$, and the scaling exponents now read $\theta= 4 \pm 6.18i$.  
Overall, the RG flow is topologically equivalent to \eq{beta}, \eq{Landau}.
We  conclude that  the qualitative difference between Fig.~\ref{pMS} and Fig.~\ref{p4d} is  due to the effective dimensionality in which the conformal factor fluctuates, given by the parameter $n$.

\begin{figure}[t]
\begin{center}
\unitlength0.001\hsize
\begin{picture}(1000,620)
\put(20,0){\includegraphics[scale=.75]{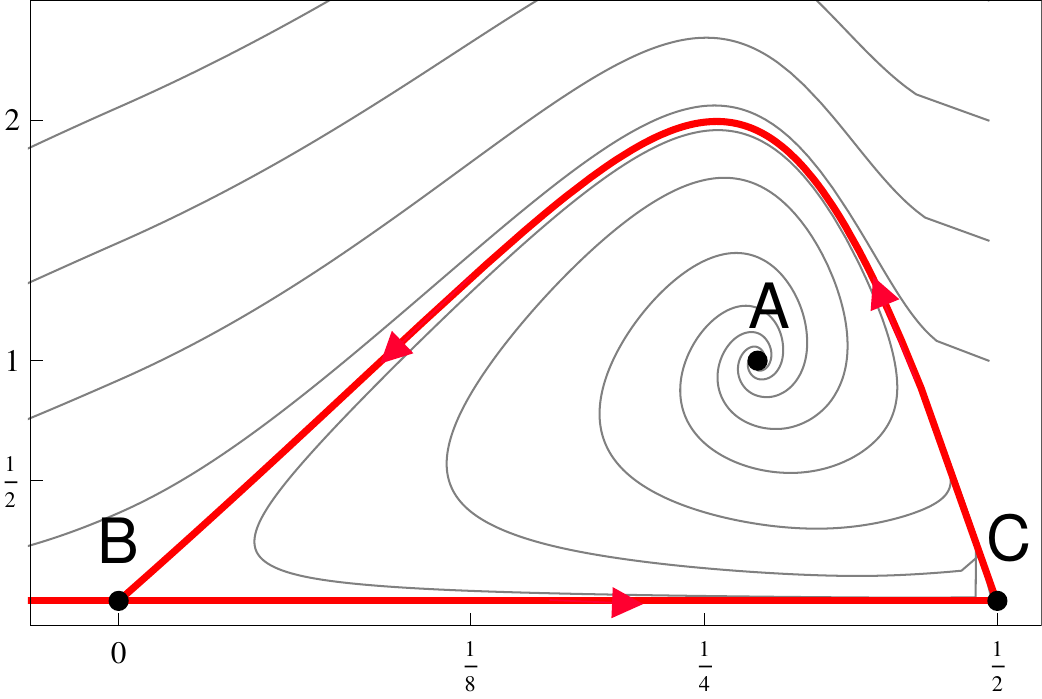}}
\put(-20,560){\Large$\frac{g}{g_*}$}
\put(940,10){\Large $\lambda$}
\end{picture}
\caption{Fixed points and degenerate limit cycle ($n=n_{\rm crit}$).}
\vskip-.7cm
\label{pEff}
\end{center}\end{figure}

To interpolate between  theses cases
we consider a model where
$n=n_{\rm eff}$ is treated as an effective, adjustable
parameter. 
A further motivation for this is that gravity appears to become lower-dimensional in response to its own quantum dynamics  \cite{Litim:2006dx,Carlip:2009km},
a behavior which can be mimicked by our model for $n_{\rm eff}<4$. Using \eq{beta} with \eq{aMS}, and generic $n_{\rm eff}$,
we find that the four fixed points $A, B, C$ and $D$ persist. In addition, we find a critical point
\begin{equation}\label{ncrit}
n_{\rm crit}=1.4715\cdots
\end{equation}
at which the phase diagram bifurcates from Fig.~\ref{pMS} into Fig.~\ref{p4d} (see Fig.~\ref{pEff}).  Approaching $n_{\rm crit}$ from below the period of the limit cycle diverges, $1/T\to 0$.  Exactly at \eq{ncrit},
the limit cycle becomes degenerate and turns into a  separatrix connecting the fixed point $C$ with the Gaussian fixed point $B$. Therefore, all trajectories emanating from the UV fixed point approach the Gaussian fixed point arbitrarily close and display  long semi-classical regimes. Furthermore, none of the asymptotically safe trajectories admits a negative cosmological constant because the fixed point $D$ is shielded from $A$ due to the IR attractor line $BC$. As soon as $n_{\rm eff}>n_{\rm crit}$, the limit cycle decays into an IR attractor line $CD$ and the separatrix  $AB$, thereby opening the door for trajectories with a cosmological constant of either sign in the IR limit. 

Finally, we  speculate on a possible link between our results and dynamical triangulations (DTs) on the lattice, assuming that the latter can be mapped
on our model for an effective parameter $n=n_{\rm lattice}$. In causal DTs, trajectories are found where
the spectral dimension 
interpolates from $D_s=4$ down to 
$D_s\approx1.80$ 
\cite{Ambjorn:2005db}. Causal DTs also show an UV fixed point \cite{Ambjorn:2011cg},
implying $n_{\rm CDT}\ge n_{\rm crit}$. 
In  euclidean DTs, the short-distance spectral dimension comes out smaller, 
$D_s\approx1.457$ 
 \cite{Laiho:2011ya}, closer to $n_{\rm crit}$, showing that causal and euclidean DTs  differ, albeit mildly, in their short-distance behavior. We therefore expect that $n_{\rm EDT}< n_{\rm CDT}$. If $ n_{\rm EDT}< n_{\rm crit}$ our  model states that euclidean DTs cannot reach an UV fixed point, in agreement with \cite{Bialas:1996wu}.
 In turn, for euclidean DTs  where  $n_{\rm EDT}$ comes out larger than \eq{ncrit},  the underlying theory can become asymptotically safe in its own right.

 In summary, we have studied models
of $4d$ quantum gravity 
with an asymptotically safe UV fixed point, and  IR fixed points corresponding to general relativity with negative, vanishing, or positive cosmological constant. 
 Interestingly, the RG flow may shield itself from asymptotic safety
by means of a  limit cycle,
which arises through competing fixed points. Roughly, the smaller the product of ultraviolet parameters $D\cdot T$, \eq{LC}, the more likely is the occurrence of an infrared limit cycle. In our model, this is governed  via the parameter $n$, which controls the short-distance fluctuations as seen in the minisuperspace.
 This pattern may also occur in  $f(R)$-theories of gravity, possibly coupled to matter,
as these share key features with the models studied here \cite{Codello:2008vh}. 
Predictions of our model can also be tested on the lattice, which is sensitive to short-distance effects. Finally we mention that  a degenerate  limit cycle offers a new scenario for asymptotic safety in that  no fine-tuning of UV initial conditions is required. Instead, all trajectories enter an extended semiclassical regime and lead to  a tiny positive cosmological constant in the infrared.  

 {\it Acknowledgements.---} 
{We thank John Cardy and Christoph Rahmede for discussions. AS thanks the University of Sussex for hospitality. 
 This work is support by CONICET  and the Science Technology and Facilities Council (STFC) under grant number ST/J000477/1.}

\end{document}